\documentclass[runningheads]{llncs}
\usepackage{graphicx}
\usepackage{float}
\usepackage{csquotes}
\usepackage{comment}
\usepackage{tikz}
\begin{document}
\title{Integrating Virtual and Augmented Reality into Public Education: Opportunities and Challenges in Language Learning
}
\titlerunning{VR and AR for Language Learning in Public Education}
%
%
\author{Tanja Kojić\inst{1} \and Maurizio Vergari\inst{1}\and Giulia-Marielena Benta\inst{1} \and Joy Krupinski\inst{1} \and Maximilian Warsinke\inst{1} \and Sebastian Möller\inst{1,3} \and Jan-Niklas Voigt-Antons\inst{2}}
\authorrunning{T. Kojić et al.}

%
\institute{Quality and Usability Lab, TU Berlin, Berlin, Germany \and Immersive Reality Lab, Hamm-Lippstadt University of Applied Sciences, Lippstadt, Germany \and German Research Center for Artificial Intelligence (DFKI), Berlin, Germany}
\maketitle    

\begin{abstract}

Virtual Reality (VR) and Augmented Reality (AR) are emerging as transformative tools in education, offering new possibilities for engagement and immersion. This paper explores their potential in language learning within public education, focusing on their ability to enhance traditional schooling methods and address existing educational gaps. The integration of VR and AR in schools, however, is not without challenges, including usability, technical barriers, and the alignment of these technologies with existing curricula. Drawing on two empirical studies, this work investigates the opportunities and challenges of VR and AR-assisted language learning, proposing strategies for their effective implementation in the public sector.
Based on two empirical studies, findings show that VR boosts motivation and immersion but has an unclear impact on vocabulary retention, with technical limitations and cognitive overload as challenges. AR enhances contextual learning and accessibility but faces usability constraints and limited personalization.
To facilitate effective adoption, this paper recommends improving interface design, reducing cognitive load, increasing adaptability, and ensuring adequate infrastructure and teacher training. Overcoming these barriers will enable a more effective integration of immersive technologies in language education.

\end{abstract}

\newcommand\copyrighttext{%
    \footnotesize \textcopyright 2026 IEEE. Personal use
    of this material is permitted. Permission from IEEE
    must be obtained for all other uses, in any current or
    future media, including reprinting/republishing this
    material for advertising or promotional purposes,
    creating new collective works, for resale or
    redistribution to servers or lists, or reuse of any
    copyrighted component of this work in other works.
    https://doi.org/10.1007/978-3-031-93715-6\_9}

\newcommand\copyrightnotice{%
\begin{tikzpicture}[remember picture,overlay,shift=
    {(current page.south)}]
    \node[anchor=south,yshift=10pt] at (0,0)
    {\fbox{\parbox{\dimexpr\textwidth-\fboxsep-
    \fboxrule\relax}{\copyrighttext}}};
\end{tikzpicture}%
}
\copyrightnotice

%
%
%









\setcounter{footnote}{0} 

\section{Introduction}

Virtual Reality (VR) and Augmented Reality (AR) are becoming popular tools for learning languages. These technologies create interactive and immersive experiences, making learning more engaging. Traditional language learning often involves memorization and passive study, but VR and AR allow learners to interact with realistic situations. These experiences can increase motivation and improve learning \cite{deng2022}.

VR places learners in fully immersive environments where they can practice language skills in lifelike scenarios. Research shows that VR makes learning more exciting and engaging. However, its impact on language improvement, especially vocabulary learning, is not clear \cite{nicolaidou2021}. Some studies suggest that while learners enjoy VR, the complexity of using it can make learning harder. Also, the need for special devices limits its use in schools \cite{parmaxi2020}.

AR adds digital content to the real world, making learning more interactive. Unlike VR, AR works on regular mobile devices, making it easier to use. AR apps like Mondly AR help learners interact with digital elements in their surroundings, making learning more engaging and practical \cite{cai2022}. However, AR also has challenges, such as too much information at once, difficult interfaces, and lack of personalization. Making AR fit into school lessons while keeping students interested is still a challenge for developers and teachers \cite{pillai2013}.

Research shows that VR and AR both have advantages and limitations in language learning. VR helps learners feel present and involved, while AR makes learning easier and more relevant to real life \cite{schubert2001}. However, personal factors such as past experience with VR, ability to learn languages, and the number of languages a person knows can affect how useful these tools are \cite{kang2021}. It is also important to design user-friendly interfaces and effective learning methods to make these technologies work well in education \cite{hardison2012}.

The aim of this paper is to analyze the benefits and challenges of VR and AR in language learning by synthesizing insights from empirical studies and usability assessments. Specifically, the paper explores the impact of VR and AR on learner engagement and comprehension, identifies usability challenges and technological limitations, and proposes strategies for improving the effectiveness of immersive language learning tools in educational settings.

\section{Related Work}
\label{sec:related}

\subsection{Augmented Reality in Language Learning}

Augmented Reality has gained attention in language learning for its ability to merge virtual elements with real-world environments, fostering interactive and engaging experiences. Studies highlight AR’s potential to improve comprehension and retention through multimodal interaction, particularly in vocabulary acquisition, pronunciation, and reading comprehension. Despite these advantages, its integration into educational frameworks remains limited, necessitating further research on long-term effectiveness and adaptability.


Research indicates that AR enhances learning outcomes by providing personalized instruction, interactive feedback, and gamification elements. The ability to manipulate virtual objects and engage in real-time interactions strengthens motivation and reinforces learning. Features such as adaptive content, which adjusts based on user proficiency, allow for a more tailored educational experience. Gamification, including rewards and interactive challenges, has also been shown to sustain learner engagement over time.

Although AR has been successfully implemented in vocabulary-building and pronunciation exercises, its application to higher-order language skills remains underexplored. Writing, grammar acquisition, and discourse-based learning still require significant development in AR-based methodologies.


Despite its potential, several barriers hinder the widespread adoption of AR in language education. Technical limitations, such as high development costs, hardware constraints, and software compatibility, impact accessibility. Additionally, the cognitive demands of AR environments, where multiple sensory inputs must be processed simultaneously, can lead to information overload, reducing learning efficiency. Another challenge lies in usability, where poorly designed interfaces and unclear navigation reduce the effectiveness of AR-based learning tools. While AR demonstrates strong potential in foundational language skills, its capacity to support advanced linguistic competencies, including writing and critical thinking, remains an area requiring further investigation.


Research suggests that usability plays a crucial role in determining the success of AR-based language learning. Factors such as intuitive navigation, interactive storytelling, and real-time feedback mechanisms contribute significantly to engagement and learning retention. Adaptive learning environments, where content adjusts based on user progress, have been shown to enhance user satisfaction and improve language acquisition \cite{wedyan2022augmented}. 
Beyond usability, learner background and educational setting influence AR’s effectiveness \cite{alvarez2024insights}. Prior exposure to AR, language proficiency levels, and learning context—whether self-directed or classroom-based—impact how users engage with and benefit from AR applications. Addressing these factors through customized experiences can improve accessibility and maximize AR’s educational potential.


To overcome existing challenges, researchers propose advancements in AI-driven personalization, enabling content to dynamically adapt to individual learning styles and progress \cite{imran2024personalization}. Enhanced gamification techniques, incorporating interactive elements and real-time feedback, can further sustain motivation and engagement. The development of cloud-based and cross-device platforms is also recommended to improve accessibility, ensuring seamless integration across different learning environments. Moreover, structured teacher training programs and pedagogical support will be essential in bridging the gap between AR technology and effective educational implementation. While AR holds significant promise, future studies should explore its role in more complex linguistic tasks, refine usability strategies, and develop scalable learning models for broader adoption.

\subsection{Virtual Reality in Language Learning}

Virtual Reality has emerged as an immersive alternative for language learning, offering real-time simulated interactions and contextualized learning experiences. Studies suggest that VR can enhance listening, pronunciation, and conversational fluency, but its effectiveness in reading comprehension, vocabulary retention, and writing remains debated \cite{deng2022}.


Language acquisition is shaped by various cognitive and environmental factors. Age plays a significant role, with younger learners relying more on intuitive absorption, whereas adults depend on analytical reasoning and structured learning \cite{dekeyser2000robustness}. Language aptitude, particularly associative memory and phonological short-term memory, has been identified as a key predictor of language learning success \cite{wen2017latent}. Additionally, immersion remains a crucial component, as extended exposure to a language environment strengthens neural pathways associated with language processing. Another important factor is cross-linguistic transfer, where prior knowledge of a second language facilitates the acquisition of a third language, demonstrating the cognitive benefits of multilingual learning \cite{kang2021}.


A systematic review examined VR-assisted language learning (VRALL) research from 2015 to 2018, finding that VR enhances engagement and motivation but lacks extensive studies on long-term learning outcomes \cite{parmaxi2020}. More recent work, which analyzed studies from 2018 to 2022, noted a significant increase in VR adoption, particularly following the COVID-19 pandemic. While VR was found to improve learner confidence and immersion, concerns were raised about content authenticity and the cognitive load associated with virtual environments \cite{hua2023}.
Research indicates that VR is particularly effective in developing pronunciation and conversational fluency, whereas vocabulary acquisition, reading comprehension, and writing proficiency show more mixed results. The lack of standardized learning models in VR-based education further complicates its integration into formal curricula.


This study builds on previous research, which investigated the effectiveness of the Mondly VR language-learning application. Their findings indicated that while VR increased engagement and immersion, it did not result in significantly higher language competence compared to mobile learning \cite{nicolaidou2021}. Expanding on these insights, the present study explores additional variables such as device variability and the role of third language acquisition in VR-assisted learning. By incorporating cognitive and pedagogical factors, this research aims to deepen understanding of VR’s educational potential.


To enhance the efficacy of VR in language education, researchers emphasize the need for improved content authenticity, ensuring culturally relevant and realistic interactions. The integration of AI-driven adaptive learning models could provide personalized instruction, dynamically adjusting content based on learner progress. Furthermore, cross-platform compatibility between VR, mobile, and desktop applications would facilitate flexible learning pathways. The alignment of VR technologies with structured curricula and teacher training programs is also crucial to maximizing its educational benefits. While VR has demonstrated clear advantages in fostering immersion and engagement, future research must focus on evaluating its effectiveness across a wider range of linguistic skills, refining interactive methodologies, and ensuring scalability for diverse educational contexts.

\subsection{Comparison of AR and VR in Language Learning}

Both AR and VR offer different benefits for language learning, each supporting different parts of the learning process. AR adds digital content to real-world settings, making it useful for vocabulary learning and interactive practice. Gamification and real-world context help learners stay engaged, especially beginners.
VR, on the other hand, creates immersive environments that help with pronunciation, listening comprehension, and conversation skills. By simulating real-life situations, VR allows learners to practice language in a natural way.

Despite these advantages, both technologies have challenges in education. Technical issues, usability problems, and the need for more adaptable content make it difficult to integrate them into schools. To make AR and VR more effective, future research should focus on improving user-friendly designs, expanding learning models beyond basic skills, and studying their long-term impact on language learning. Overcoming these challenges will help ensure that AR and VR are not just engaging but also truly beneficial for language education.

\section{Methodology}

This section describes the research approach used to investigate the role of AR and VR in language learning. Two separate empirical studies were conducted to assess the impact of these immersive technologies on different aspects of language acquisition. The first study examined the usability and learning experience in an AR-based language learning application, while the second study evaluated how VR immersion influences pronunciation, listening comprehension, and conversational practice. A mixed-method approach was used in both studies, combining quantitative learning assessments with qualitative feedback analysis to provide a comprehensive understanding of their effectiveness.

\subsection{AR Study: Usability and Learning Experience}
The AR study focused on evaluating the role of Mondly AR, a widely used augmented reality language learning application. The research assessed usability, learner engagement, and cognitive load while exploring how interactive AR elements influence vocabulary acquisition.

\subsubsection{Participants and Study Design}
Participants were recruited through academic networks and online advertisements, targeting individuals with varying levels of language proficiency. The study aimed to include both novice and experienced users of AR technology to understand its accessibility across different learner backgrounds.
A controlled experimental design ensured that all participants interacted with the AR application under standardized conditions. Each participant engaged in a structured learning session, allowing for a consistent evaluation of usability and learning outcomes.

\subsubsection{Application}
The Mondly AR application was selected for its ability to overlay virtual language-learning elements onto real-world environments. The app offers interactive vocabulary exercises, pronunciation feedback, and conversational practice using speech recognition. Learners engage with 3D-rendered objects that represent words in the target language, allowing for contextualized language acquisition. The app also incorporates gamification elements, such as progress tracking and achievement rewards, to maintain learner motivation.
Figure \ref{fig:mondly_vr} illustrates a typical scenario, where users engage in real-time spoken interactions within a simulated environment.
The study assessed the effectiveness of these features in supporting vocabulary retention while also identifying potential usability challenges. Particular attention was given to interface design, ease of navigation, and cognitive load, as AR applications require learners to process both virtual and real-world stimuli simultaneously.

\subsubsection{Procedure}
Each participant completed a three-phase structured session, beginning with a pre-test to assess their initial vocabulary knowledge. They then engaged with the Mondly AR application, interacting with augmented objects and practicing pronunciation through speech recognition exercises. The session concluded with a post-test measuring vocabulary retention and a usability questionnaire evaluating the overall learning experience.
%

\subsubsection{Data Collection and Analysis}
Data was collected through a combination of learning assessments, usability questionnaires, and qualitative feedback. The pre- and post-test scores provided a measure of vocabulary improvement, while usability ratings captured participants’ perceptions of ease of use, responsiveness, and interface clarity.
Qualitative responses were analyzed to gain insights into learner adaptation, cognitive demands, and overall satisfaction with AR-based language learning. Thematic analysis was used to identify recurring usability challenges and determine how well AR supports interactive language acquisition.

\subsection{VR Study: Immersion and Language Acquisition}
The VR study investigated the role of ImmerseMe VR, an immersive language-learning application designed for conversational practice and pronunciation training. The study explored how virtual environments influence learner engagement and whether VR enhances language learning beyond traditional methods.

\subsubsection{Participants and Study Design}
Similar to the AR study, participants were recruited through academic networks and online platforms, ensuring a diverse sample with varying degrees of VR experience. The study followed a between-subjects design, where one group completed language exercises in VR while a control group used the same application on a standard desktop interface.
The inclusion of both VR and non-VR conditions allowed for a direct comparison between immersive and traditional learning methods, providing insights into the added value of virtual environments.

\subsubsection{Application}
\begin{figure}[H]
 \centering
 \begin{minipage}{0.49\textwidth}
  \centering
  \includegraphics[height=6cm]{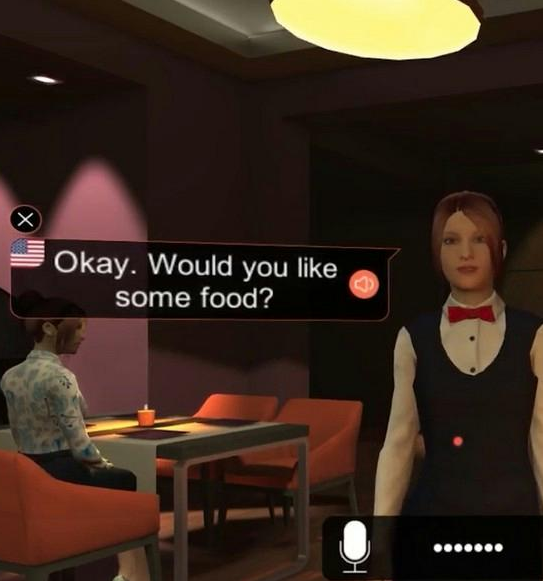}
  \caption{A screenshot of a conversational Mondly VR scenario.}
  \label{fig:mondly_vr}
 \end{minipage}
 \hfill
 \begin{minipage}{0.49\textwidth}
  \centering
  \includegraphics[height=6cm]{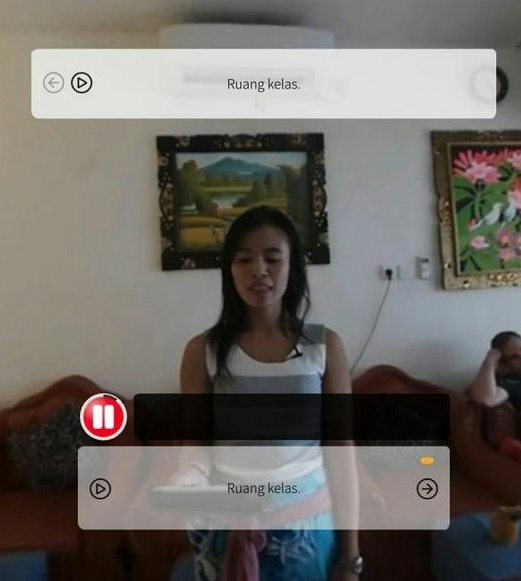}
  \caption{A screenshot of the vocabulary module in ImmerseMe.}
  \label{fig:immerseme_vocabulary}
 \end{minipage}
\end{figure}

The ImmerseMe VR application was selected due to its focus on simulated real-world conversations and interactive speech-based exercises. 
As shown in Figure \ref{fig:immerseme_vocabulary}, learners engage with interactive word-learning exercises, reinforcing language retention through real-time feedback.
Unlike traditional language-learning software, ImmerseMe VR places learners in culturally relevant environments, such as restaurants, airports, and markets, where they must navigate conversations naturally. The app's adaptive difficulty levels allow learners to progress based on their speech accuracy and fluency.
The study examined how these immersive elements affected user engagement and whether they contributed to improvements in pronunciation and conversational confidence. It also assessed usability factors, particularly the ease of interaction within VR environments, responsiveness of the speech recognition system, and potential challenges such as motion discomfort.

\subsubsection{Procedure}
Each participant completed a structured learning session consisting of an introduction, pre-test, interaction phase, and post-test. At the start, they were introduced to the VR headset and the ImmerseMe application, followed by a pre-test evaluating their pronunciation and listening comprehension skills.

Participants were assigned to different experimental conditions based on the application, device, and target language. The study included four conditions for each application (Mondly and ImmerseMe), resulting in a total of eight experimental groups. Table \ref{tab:experiment_conditions} outlines the distribution of conditions.

\begin{table}[H]
\vspace{-1em}
 \centering
 \caption{The experiment conditions}

 \label{tab:experiment_conditions}
 \begin{tabular}{|l|l|l|l|}
  \hline
  \textbf{Application} & \textbf{Device} & \textbf{Language} & \textbf{Code} \\ 
  \hline
  Mondly VR & VR Headset & Greek & C1 \\ 
  Mondly VR & VR Headset & Indonesian & C2 \\ 
  Mondly PC & Desktop & Greek & C3 \\ 
  Mondly PC & Desktop & Indonesian & C4 \\ 
  ImmerseMe VR & VR Headset & Greek & C5 \\ 
  ImmerseMe VR & VR Headset & Indonesian & C6 \\ 
  ImmerseMe PC & Desktop & Greek & C7 \\ 
  ImmerseMe PC & Desktop & Indonesian & C8 \\ 
  \hline
 \end{tabular}
 \vspace{-1em}
\end{table}

Following the pre-test, participants engaged with either Mondly VR or ImmerseMe VR using a virtual reality headset or completed the same exercises on a desktop version of the application. Those in the VR groups practiced interactive dialogues and pronunciation exercises within immersive scenarios, while participants in the PC-based conditions followed structured text and audio prompts.

After the interaction phase, all participants completed a post-test to measure pronunciation improvements and comprehension accuracy. Those in the VR conditions additionally completed the Igroup Presence Questionnaire (IPQ) to assess their sense of presence in the virtual environment. A usability questionnaire was also administered to evaluate navigation, cognitive load, and the overall learning experience.

\subsubsection{Data Collection and Analysis}
A mixed-method approach was used to analyze learning performance, engagement, and usability factors. Pronunciation and comprehension scores from pre- and post-tests were compared across VR and non-VR conditions to determine the effectiveness of immersion in language learning.
In addition to quantitative assessments, qualitative feedback was collected through open-ended user reflections, where participants described their experiences, challenges, and perceived benefits of VR-based learning. This feedback was analyzed to identify common themes, such as learner motivation, usability challenges, and cognitive workload.
By comparing the VR and non-VR groups, the study provided insights into how immersion influences language retention and whether virtual scenarios enhance engagement and confidence in conversational practice.

\subsection{Ethical Considerations}
Both studies were conducted in accordance with ethical research guidelines, ensuring participant safety, privacy, and informed consent. Prior to participation, all individuals received a detailed explanation of the study objectives, procedures, and potential risks. Participants voluntarily provided informed consent and were given the option to withdraw from the study at any stage without consequences.
All collected data was anonymized, and no personally identifiable information was recorded. The research followed data protection regulations, ensuring secure storage and responsible handling of participant responses. By maintaining these ethical standards, the study ensured the reliability and validity of its findings while prioritizing participant well-being.

\section{Results}

This section presents the findings from the AR and VR studies, focusing on the impact of both technologies on vocabulary acquisition, pronunciation, usability, and user experience. The results are structured to allow for a clear comparison of both immersive learning technologies while also highlighting individual strengths and limitations. The analysis is based on both quantitative performance metrics and qualitative user feedback.

\subsection{AR Study Results}

The AR study aimed to examine how augmented reality supports vocabulary learning, particularly in terms of usability and learner engagement. The study included 45 participants, each engaging with Mondly AR for structured vocabulary training. To assess the effectiveness of AR-assisted language learning, a pre-test and post-test comparison was conducted, revealing a statistically significant improvement of 12.5\% in vocabulary acquisition (p = 0.03). This suggests that AR-based learning can enhance language retention in the short term.

\begin{figure}[H]
    \centering
    \includegraphics[width=0.7\textwidth]{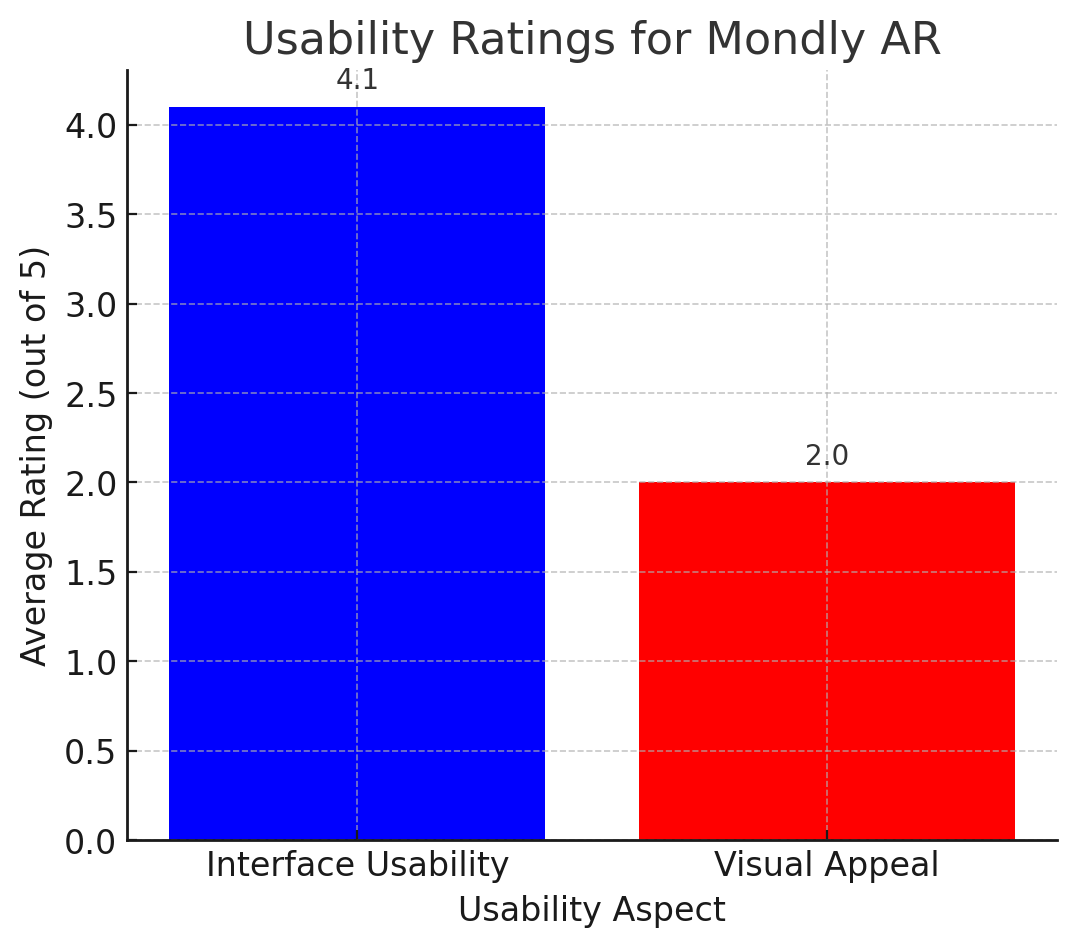}
    \caption{Usability Ratings for Mondly AR, comparing interface usability and visual appeal.}
    \label{fig:usability_mondly_ar}
\end{figure}

Participants were also asked to evaluate the usability of the Mondly AR application (restuls shown with Figure \ref{fig:usability_mondly_ar}, which received an average rating of 4.1 out of 5, indicating high user satisfaction with the interface design and navigation. Eighty percent of participants found the app easy to navigate, highlighting its accessibility for learners with varying levels of technological experience. However, fifty percent of users rated the visual appeal as only 2 out of 5, suggesting that while the interface was functional, improvements could be made to enhance the visual design.

A key aspect of the study was understanding the relationship between usability and learning performance. Correlation analysis revealed a positive relationship between usability ratings and vocabulary gains (r = 0.42, p = 0.02). This finding indicates that a well-designed interface and seamless user experience contribute to better learning outcomes. However, despite improvements in vocabulary retention, seventy percent of participants reported no significant increase in their confidence in speaking the target language after using the application. This suggests that while AR is effective for vocabulary acquisition, it may not sufficiently support broader language skills such as fluency and spontaneous speech production.

Another noteworthy aspect was the demand for greater customization in AR learning environments. Sixty-six point seven percent of users expressed a desire for adjustable settings that would allow them to tailor the difficulty level, interaction type, or pace of learning. This highlights the importance of personalization in educational applications to accommodate different learning preferences.

The AR study demonstrated that augmented reality can be an effective tool for vocabulary acquisition, offering an engaging and interactive way to reinforce learning. However, limitations such as low visual appeal, limited long-term impact on confidence, and the lack of personalization suggest areas for future improvement.

\subsection{VR Study Results}

The VR study focused on pronunciation training, listening comprehension, and conversational practice, comparing the effectiveness of VR-based learning to non-VR methods. A total of 31 participants were divided into two conditions: a VR group that engaged with ImmerseMe VR and a non-VR control group that used a desktop version of the same application. The goal was to determine whether immersion in a virtual environment leads to higher learning gains compared to traditional screen-based learning.

Performance assessments showed that participants in the VR condition demonstrated an average improvement of 8.3 points (SD = 2.1) in their language test scores, whereas those in the non-VR condition achieved an average gain of 4.5 points (SD = 1.8). This indicates that VR-based language learning provides a measurable advantage over non-immersive methods, supporting previous research suggesting that immersion enhances retention and engagement.

\begin{figure}[H]
    \centering
    \includegraphics[width=0.7\textwidth]{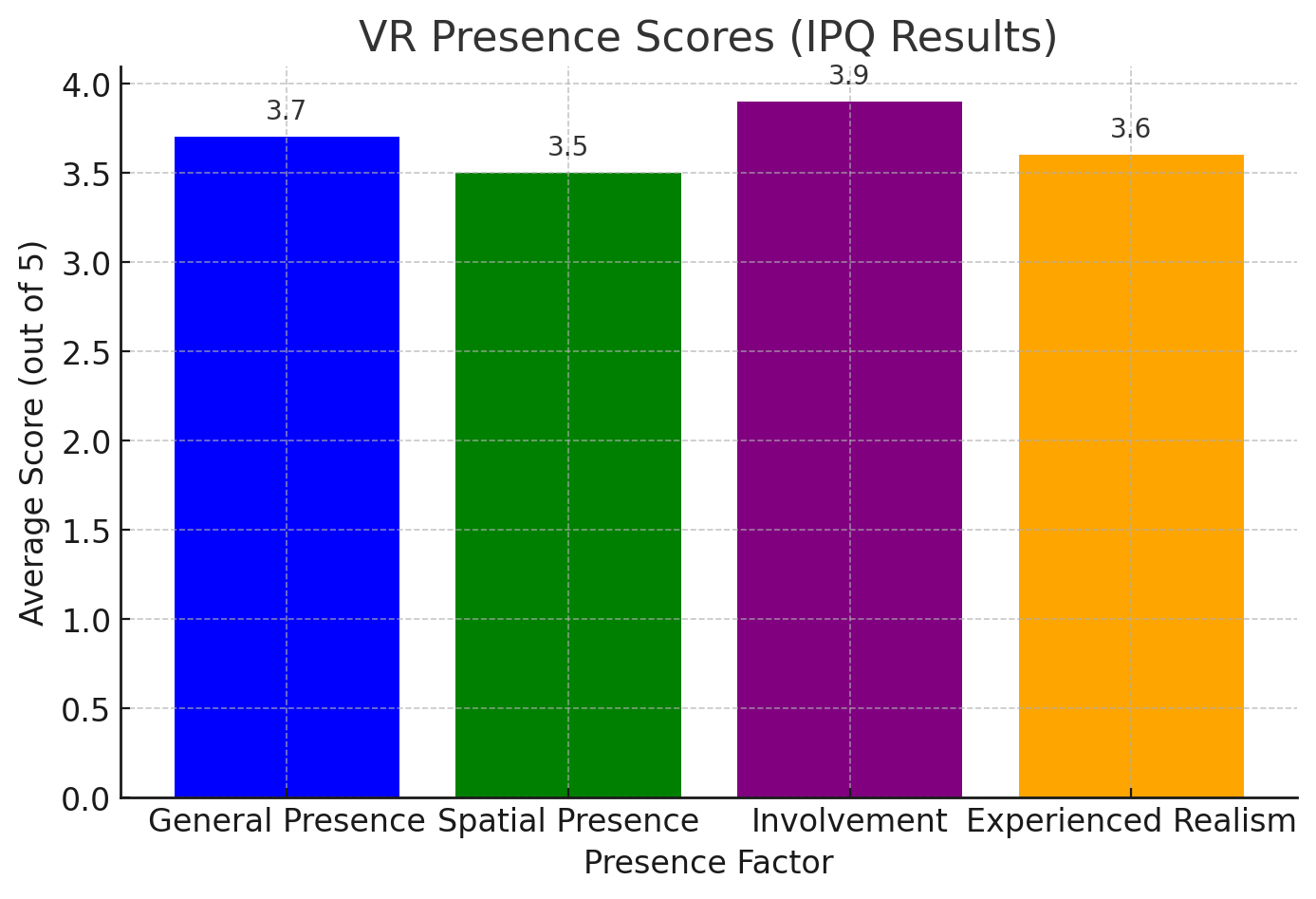}
    \caption{VR Presence Scores (IPQ results), showing General Presence, Spatial Presence, Involvement, and Experienced Realism.}
    \label{fig:vr_presence_scores}
\end{figure}

To further investigate the role of immersion, the Igroup Presence Questionnaire (IPQ) was used to assess the perceived level of presence within the virtual environment, as shown with Figure \ref{fig:vr_presence_scores}. The results for the VR group were as follows: general presence score of 3.7, spatial presence score of 3.5, involvement score of 3.9, and experienced realism score of 3.6. These scores indicate a moderate-to-high level of presence, suggesting that participants felt relatively immersed in the virtual environment. Notably, involvement received the highest score, which may reflect the interactive nature of VR-based language tasks.

Session duration was also analyzed, with VR learners spending an average of 7.32 minutes per round (SD = 1.74 min), compared to 6.03 minutes (SD = 2.16 min) in non-VR conditions. This suggests that VR environments encourage longer and more engaged interactions compared to desktop-based learning.

Further analysis examined the correlation between presence and learning gains. A moderate positive correlation (r = 0.45, p = 0.04) was found between overall presence ratings and improvements in language scores. This suggests that the more immersed participants felt in the VR environment, the more they benefited from the learning experience.

Session duration varied between applications. Participants in Mondly VR spent an average of 7.54 minutes, while those using ImmerseMe VR engaged for 7.10 minutes per session. In contrast, non-VR users had significantly shorter interactions, with Mondly PC averaging 7.08 minutes and ImmerseMe PC averaging only 4.98 minutes. These findings indicate that VR encourages longer engagement, which may contribute to better learning retention.

Despite these positive findings, no significant correlation was found between prior VR experience and language improvement. This suggests that VR-based learning can be effective for both experienced and novice users, making it accessible to a wider range of learners. While VR appears to be a promising tool for language learning, future research should explore how different levels of interactivity and feedback mechanisms influence the long-term retention of language skills.

\section{Discussion}

This section discusses the key findings of the study, comparing the impact of AR and VR on language learning while considering their strengths, limitations, and alignment with previous research. The implications of these results for educational practice and future research are also explored.

\subsection{Comparison of AR and VR Findings}

The results indicate that both AR and VR enhance language learning, but they serve different purposes and impact learners in distinct ways. The AR study demonstrated that augmented reality is effective for structured vocabulary learning, supporting previous research that highlights AR's ability to integrate digital content into real-world settings and reinforce word associations through interactive engagement \cite{cai2022}. The usability ratings suggest that AR applications are generally accessible and intuitive, making them suitable for a wide range of learners. However, despite positive usability feedback, AR was found to have limited influence on spoken language confidence, which aligns with prior findings that AR-based learning tends to focus on object recognition and passive vocabulary recall rather than conversational fluency \cite{nicolaidou2021}.

In contrast, VR was found to be more effective in supporting pronunciation and conversational practice, aligning with research that emphasizes the role of immersion and presence in second language acquisition \cite{schubert2001}. Participants reported high levels of engagement and involvement, suggesting that VR fosters a more interactive and dynamic learning environment compared to traditional or screen-based learning methods. These findings are consistent with studies that highlight the role of spatial presence and realism in language immersion, where learners engage in simulated conversations that closely mimic real-world interactions \cite{kang2021}.

A key distinction between the two technologies is how they influence learning engagement and cognitive load. AR applications provided structured learning experiences that were easier to navigate and adapt to, but they lacked customization features that could enhance long-term engagement. Previous studies have noted similar findings, where AR is effective for controlled learning tasks but does not inherently encourage spontaneous or adaptive learning behaviors \cite{parmaxi2020}. In contrast, VR-based learning required longer session durations, which may indicate a higher cognitive demand associated with navigating virtual spaces, processing interactive dialogues, and maintaining presence \cite{hua2023}. While the increased engagement in VR can be beneficial, it may also lead to higher cognitive load, making it necessary to optimize session length and interaction complexity for different learner levels.

The differences in usability ratings also reflect broader challenges in immersive learning environments. AR was rated highly for usability and interface clarity, reinforcing the idea that AR applications, particularly those designed for mobile, benefit from a familiar interface and minimal hardware requirements \cite{cai2022}. 

However, its effectiveness in language retention and spontaneous language use remains limited. VR, on the other hand, was perceived as more engaging and immersive, but it also presented accessibility barriers due to hardware costs and learning curve. This contrast aligns with previous discussions on the trade-offs between immersion and accessibility in language technology \cite{parmaxi2020}.

Taken together, these findings suggest that AR and VR support different stages of language learning. AR is well-suited for beginners focusing on vocabulary acquisition and structured exercises, while VR is more beneficial for learners looking to practice conversational fluency in immersive contexts. This differentiation is crucial for educators and developers aiming to design adaptive and hybrid learning environments that integrate both technologies effectively.

\subsection{Implications and Future Directions}

The findings of this study have important implications for educational practice, technology development, and future research. The effectiveness of AR in vocabulary acquisition suggests that it can be integrated into classroom settings as a supplement to traditional language instruction. However, the low impact on conversational skills highlights the need for adaptive learning models that incorporate dialogue-based interaction, speech recognition, and contextualized language exercises. Future AR applications should also focus on improving interface aesthetics and expanding user control over learning paths to foster better engagement.

The strong engagement and presence reported in VR learning underscore the importance of immersion in language acquisition. However, VR also presents challenges such as higher cognitive load, extended session durations, and accessibility limitations due to hardware requirements. 

To address these challenges, structured guidance, adaptive difficulty levels, and personalized feedback mechanisms should be incorporated into VR-based learning experiences \cite{hua2023}. Previous research suggests that shorter, goal-oriented VR sessions may be more effective in preventing cognitive overload while maintaining engagement \cite{parmaxi2020}.

A key insight from this study is that prior VR experience did not significantly influence learning outcomes, suggesting that VR learning can be accessible to a broad range of users, including those unfamiliar with immersive technology. However, the long-term effects of VR-based language learning remain underexplored. Future studies should examine whether repeated exposure to VR enhances language retention over time and whether combining AR and VR creates a more comprehensive learning experience \cite{cai2022}.

Additionally, future research should explore the social dimensions of immersive language learning. Studies have shown that collaborative learning in VR settings can increase motivation and knowledge retention, yet the potential for multi-user VR language learning environments remains largely untapped \cite{parmaxi2020}. Incorporating peer interaction and real-time feedback could significantly enhance the effectiveness of VR-based conversational training.

Another key consideration is infrastructure and accessibility. While VR offers strong engagement benefits, its adoption in public education remains limited by cost, space requirements, and motion sickness concerns. AR, being more accessible through standard mobile devices, provides a practical alternative for schools and educational institutions with limited budgets. Future research should investigate how to make VR learning more scalable and cost-effective, potentially through cloud-based VR solutions or simplified headset designs that reduce barriers to adoption.

Overall, this paper reinforces the idea that AR and VR each play a unique role in language learning. AR applications are effective for structured learning, particularly in vocabulary acquisition, whereas VR provides immersive, interactive experiences that enhance pronunciation and conversational fluency. Future research should focus on hybrid learning models that integrate AR for structured learning and VR for real-world interaction, maximizing the benefits of both technologies. By refining interface design, improving personalization, and optimizing session structures, immersive learning technologies can be developed into powerful tools that support diverse language learners across different learning environments.

\section{Conclusion}
Language learning benefits from both AR and VR, as each technology brings unique advantages while also presenting certain challenges. AR offers interactive and engaging learning experiences but is constrained by usability challenges and a lack of structured content. VR fosters high levels of engagement and immersion, yet its impact on language acquisition remains inconclusive. Future research should prioritize refining usability, integrating adaptive learning frameworks, and aligning immersive technologies with pedagogical best practices to maximize their educational potential.

%
%
%
\bibliographystyle{splncs04}
\bibliography{main}

@article{deng2022,
  author = {Deng, L. and Yu, S.},
  title = {The Impact of Virtual Reality on Language Learning: A Systematic Review},
  journal = {Educational Technology \& Society},
  volume = {25},
  number = {3},
  pages = {45-58},
  year = {2022}
}

@article{hua2023,
  author = {Hua, X. and Wang, J.},
  title = {Virtual Reality and Augmented Reality in Language Education: Advances and Challenges},
  journal = {Computers \& Education},
  volume = {185},
  pages = {104512},
  year = {2023},
  doi = {10.1016/j.compedu.2023.104512}
}

@article{nicolaidou2021,
  author = {Nicolaidou, I. and Antoniou, P. and Constantinou, R.},
  title = {Exploring Virtual Reality as a Medium for Second Language Acquisition: Engagement and Learning Outcomes},
  journal = {Journal of Language Learning and Technology},
  volume = {25},
  number = {2},
  pages = {30-48},
  year = {2021}
}

@article{parmaxi2020,
  author = {Parmaxi, A.},
  title = {Virtual Reality in Language Learning: A Systematic Review and Future Prospects},
  journal = {Educational Technology Research and Development},
  volume = {68},
  pages = {1509-1531},
  year = {2020},
  doi = {10.1007/s11423-020-09761-7}
}

@article{cai2022,
  author = {Cai, S. and Liu, E.},
  title = {Augmented Reality in Language Learning: Improving Engagement and Comprehension},
  journal = {International Journal of Educational Research},
  volume = {112},
  pages = {102093},
  year = {2022},
  doi = {10.1016/j.ijer.2022.102093}
}

@article{pillai2013,
  author = {Pillai, R. and Johnson, T.},
  title = {Measuring the Effects of Augmented Reality on Cognitive Load in Language Learning},
  journal = {Interactive Learning Environments},
  volume = {21},
  number = {6},
  pages = {615-633},
  year = {2013},
  doi = {10.1080/10494820.2013.832112}
}

@article{schubert2001,
  author = {Schubert, T. and Friedmann, F. and Regenbrecht, H.},
  title = {The Experience of Presence: Factor Analytic Insights},
  journal = {Presence: Teleoperators and Virtual Environments},
  volume = {10},
  number = {3},
  pages = {266-281},
  year = {2001}
}

@article{kang2021,
  author = {Kang, Y. and Shin, D.},
  title = {The Role of Prior Language Experience in Virtual Reality Language Learning},
  journal = {Second Language Research},
  volume = {37},
  number = {4},
  pages = {567-589},
  year = {2021},
  doi = {10.1177/0267658321998871}
}

@article{hardison2012,
  author = {Hardison, D. and Son, J.},
  title = {Usability and Pedagogical Design in Immersive Language Learning Environments},
  journal = {Language Learning \& Technology},
  volume = {16},
  number = {3},
  pages = {39-55},
  year = {2012}
}

@article{imran2024personalization,
  title={Personalization of E-Learning: Future Trends, Opportunities, and Challenges.},
  author={Imran, Muhammad and Almusharraf, Norah and Ahmed, Saim and Mansoor, Muhammad Ismail},
  journal={International Journal of Interactive Mobile Technologies},
  volume={18},
  number={10},
  year={2024}
}

@article{wedyan2022augmented,
  title={Augmented reality-based English language learning: importance and state of the art},
  author={Wedyan, Mohammad and Falah, Jannat and Elshaweesh, Omar and Alfalah, Salsabeel FM and Alazab, Moutaz},
  journal={Electronics},
  volume={11},
  number={17},
  pages={2692},
  year={2022},
  publisher={MDPI}
}

@article{alvarez2024insights,
  title={Insights into usability, academic outcomes, and emotional responses in an AR-Interactive learning environment},
  author={{\'A}lvarez-Mar{\'\i}n, Alejandro and Paredes-Velasco, Maximiliano and Vel{\'a}zquez-Iturbide, J {\'A}ngel and Palma-Chilla, Luis},
  journal={Interactive Learning Environments},
  pages={1--15},
  year={2024},
  publisher={Taylor \& Francis}
}

@article{dekeyser2000robustness,
  title={The robustness of critical period effects in second language acquisition},
  author={DeKeyser, Robert M},
  journal={Studies in second language acquisition},
  volume={22},
  number={4},
  pages={499--533},
  year={2000},
  publisher={Cambridge University Press}
}

@inproceedings{wen2017latent,
  title={Latent intention dialogue models},
  author={Wen, Tsung-Hsien and Miao, Yishu and Blunsom, Phil and Young, Steve},
  booktitle={International Conference on Machine Learning},
  pages={3732--3741},
  year={2017},
  organization={PMLR}
}
\end{document}